\documentclass{ptapap}
\usepackage{amsmath}
\usepackage{natbib}
\author{Ruchi Mishra}[CAMK]
\author{Miljenko \v{C}emelji\'{c}}[CAMK,IAA]
\author{W\l odek Klu\'{z}niak}[CAMK]
\affil[CAMK]{Nicolaus Copernicus Astronomical Center, Polish Academy of Sciences, Bartycka 18, 00--716 Warsaw, Poland}
\affil[IAA]{Institute of Astronomy and Astrophysics, Academia Sinica, P.O. Box
23--141, Taipei 106, Taiwan}

\title{Backflow in Accretion Disk}

\begin{document}

\maketitle

\begin{abstract}

Analytical solution for a thin hydrodynamic accretion disk shows that
for some values of the viscosity parameter part of the accretion flow
in the disk is not towards the star, but in the opposite direction. We
study thin disks by performing hydrodynamic simulations and compare the
numerical with analytical results. We confirm that for viscosity
coefficient smaller than a critical value, there is a backflow in
simulations near the disk mid-plane. The distance from the star to the
starting point of backflow is increasing with viscosity, as predicted
by the analytical solution. When the viscosity coefficient is larger
than critical, there is no backflow in the disk.
\end{abstract}

\section{Introduction}
Backflow is a flow in the mid-plane of accretion disk directed away from
central gravitating object, opposite to the direction of accretion flow.
Backflow in the accretion disk was first studied by \cite{urpina,urpinb}.
Considering average inflow velocity he obtained backflow in the mid-plane
of accretion disk for all values of viscosity parameter. \cite{kl92}
confirmed similar kind of backflow as Urpin in their numerical studies,
but only for small values of viscosity parameter. Backflow was also found
in works by \cite{roz} and \cite{igum95}.

Full three dimensional analytical solution for a thin accretion disk was
given by \citet[hereafter KK00]{KK00}. They derived the equations of a
polytropic, viscous hydrodynamical (HD) accretion disk using the Taylor
expansion in the small parameter $\epsilon=H/R$, the disk aspect ratio.
Backflow near the disk mid-plane was obtained for all the values of
viscous coefficient $\alpha <0.685$. The point where the backflow started,
stagnation radius, was found to be a function of viscosity parameter.

We shortly review the backflow in HD analytical solutions from KK00
 in \S 2. Results from our numerical simulations are presented
in \S 3, and a preliminary result in magnetic case is described in \S 4 . 

\section{Backflow in the analytical solution}
From KK00 solution, the equation for radial velocity  $V_r$ in the
equatorial plane in cylindrical coordinates is given by :\\
\begin{equation}
 V_r(r,0)= - \alpha\epsilon^{2}R \bigg(\frac{h^2}{r^{5/2}}\bigg)\bigg[2\bigg(\frac{d \ln h}{d \ln r}\bigg)-\Lambda\bigg(1+\frac{32}{15}\alpha^2\bigg)\bigg]
\end{equation}
where
\begin{equation*}
 \Lambda =\frac{11}{5}/(1+ \frac{64}{25}\alpha^2).
\end{equation*}
 
Here h is disk height, r is radial distance, $\alpha$ is the viscosity
parameter and $\Lambda$ is a function of $\alpha$.
 
 When $\alpha=0$ then $\Lambda$ reduces to 11/5 and the terms in the
square bracket of Eq.(1) becomes negative, so that $V_r(r, 0) > 0$.
Positive radial velocity indicates outflow in the equatorial plane away
from the central object. When $\alpha=1$ then $V_r(r, 0) < 0$,
which indicates that the equatorial flow is directed towards the central
object for all radii. The critical value over which there is no backflow is 
$\alpha_{cr}\approx0.685$.

The radius for which the radial velocity is zero is the starting point of
backflow. It is called the stagnation radius. The equation for stagnation
radius is given by:
\begin{equation}
\frac{r_{stag}(\alpha)}{r_+}=\frac{[1+6(\Lambda(1+\frac{32}{15}\alpha^2)-2)]^2}{[6(\Lambda(1+\frac{32}{15}\alpha^2)-2)]^2}
\end{equation}
   
A natural length scale is defined as $r_+=\Omega_{\mathrm m}^2r_{\mathrm m}^4/(GM_\star)$,
with $\Omega_{\mathrm m}$ being the Keplerian rotation rate at a distance where the
viscous torque is vanishing, $r_{\mathrm m}$.

The stagnation radius position is a function of viscous coefficient,
$\alpha$. With increasing $\alpha$, the stagnation radius is increasing.
When $\alpha$ approaches $\alpha_{\mathrm cr}$, the stagnation radius
becomes infinite, indicating there is no backflow in the disk.

\section{Backflow in numerical simulations}
\begin{figure*}
\includegraphics[width=1\columnwidth]{Av01omega2}
\includegraphics[width=1\columnwidth]{Av02omega2}
\includegraphics[width=1\columnwidth]{Av04omega2}
\includegraphics[width=1\columnwidth]{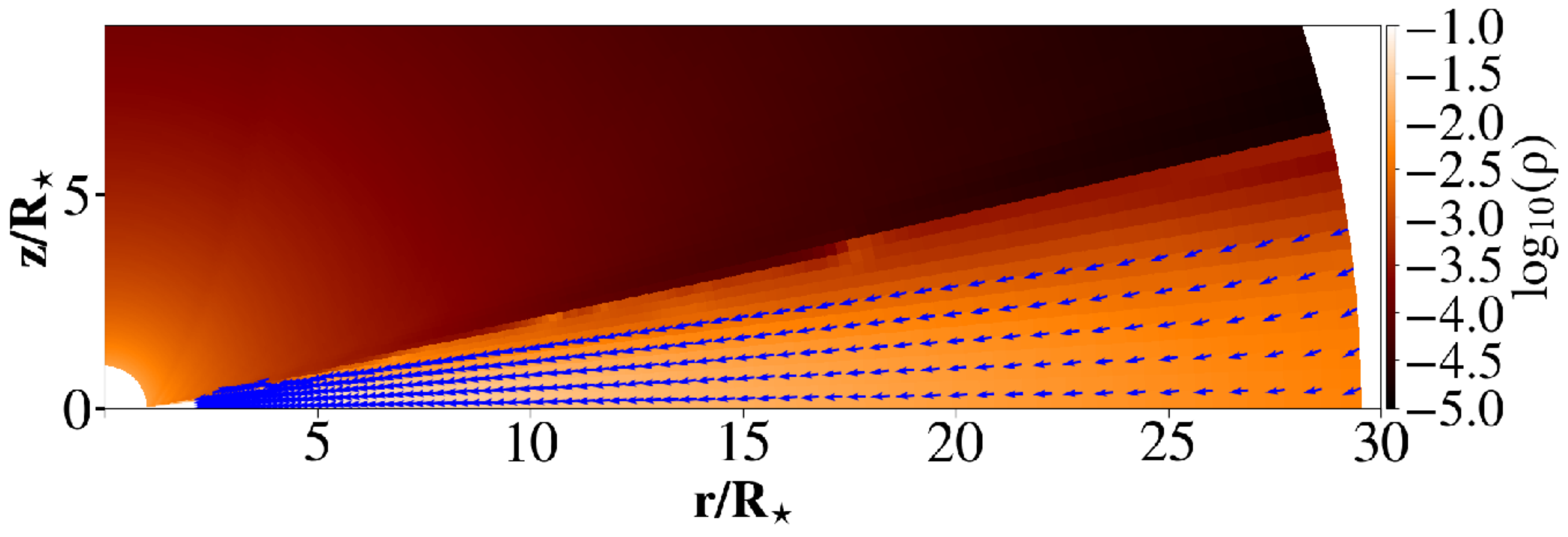}
\caption{The density in a logarithmic grading in purely HD solutions in
our simulations with $\alpha$=0.1, 0.2, 0.4 and 1, top to
bottom panels, respectively. Vectors show the velocity in the disk. The position of stagnation radius is marked $r_{stag}$.}
\label{backflshd}
\end{figure*}

We perform axisymmetric 2D star-disk simulations in viscous HD and in
resistive magneto-hydrodynamics (MHD) following \cite{zf09}. Our
setup  with the publicly available {\sc pluto} code (v.4.1)
\citep{m07,m12} is presented in detail in \cite{cem19}. A logarithmic
stretched grid in radial direction in spherical coordinates is used,
with uniformly spaced latitudinal grid. Resolution is
$R\times\theta$=[$217\times100$] grid cells, stretching the domain to
30 stellar radii(for some runs 50 stellar radii, to prevent influence
of the outer boundary) in the half of the meridional plane. Typically we
run simulations for 100 stellar rotations.

The equations solved by the {\sc pluto} code are, in the cgs units:
\begin{eqnarray}
\frac{\partial\rho}{\partial t}+\nabla\cdot(\rho\vec{\rm v}) =0\\
\frac{\partial\rho\vec{\rm v}}{\partial t}+\nabla\cdot\left[\rho\vec{\rm v}
\vec{\rm v}+\left(P+\frac{B^2}{8\pi}\right)
\vec{\rm I}-\frac{\vec{\rm B}\vec{\rm B}}{4\pi}-\vec{\tau}\right]=\rho\vec{g}\\
\frac{\partial E}{\partial t}+
\nabla\cdot\left[\left(E+P+\frac{B^2}{8\pi}\right)\vec{\rm v}-
\frac{(\vec{\rm v}\cdot\vec{\rm B})\vec{\rm B}}{4\pi}\right]\nonumber \\
+\nabla\cdot\bigg[{\eta_{\rm m}\vec{J}\times
\vec{B}/4\pi -\vec{\rm v}\cdot\vec{\tau}}\bigg]=
\rho\vec{g}\cdot\vec{\rm v}-{\Lambda} \\
\frac{\partial\vec{\rm B}}{\partial t}+\nabla\times(\vec{\rm B}
\times\vec{\rm v}+\eta_{\rm m}\vec{J})=0 
\end{eqnarray}
where the symbols have their usual meaning: $\rho$ and v are
the matter density and velocity, P is the pressure, B
is the magnetic field and $\eta_{\rm m}$ and $\tau$ represent the resistivity
and the viscous stress tensor, respectively.

\begin{figure}
\includegraphics[width=0.6\columnwidth]{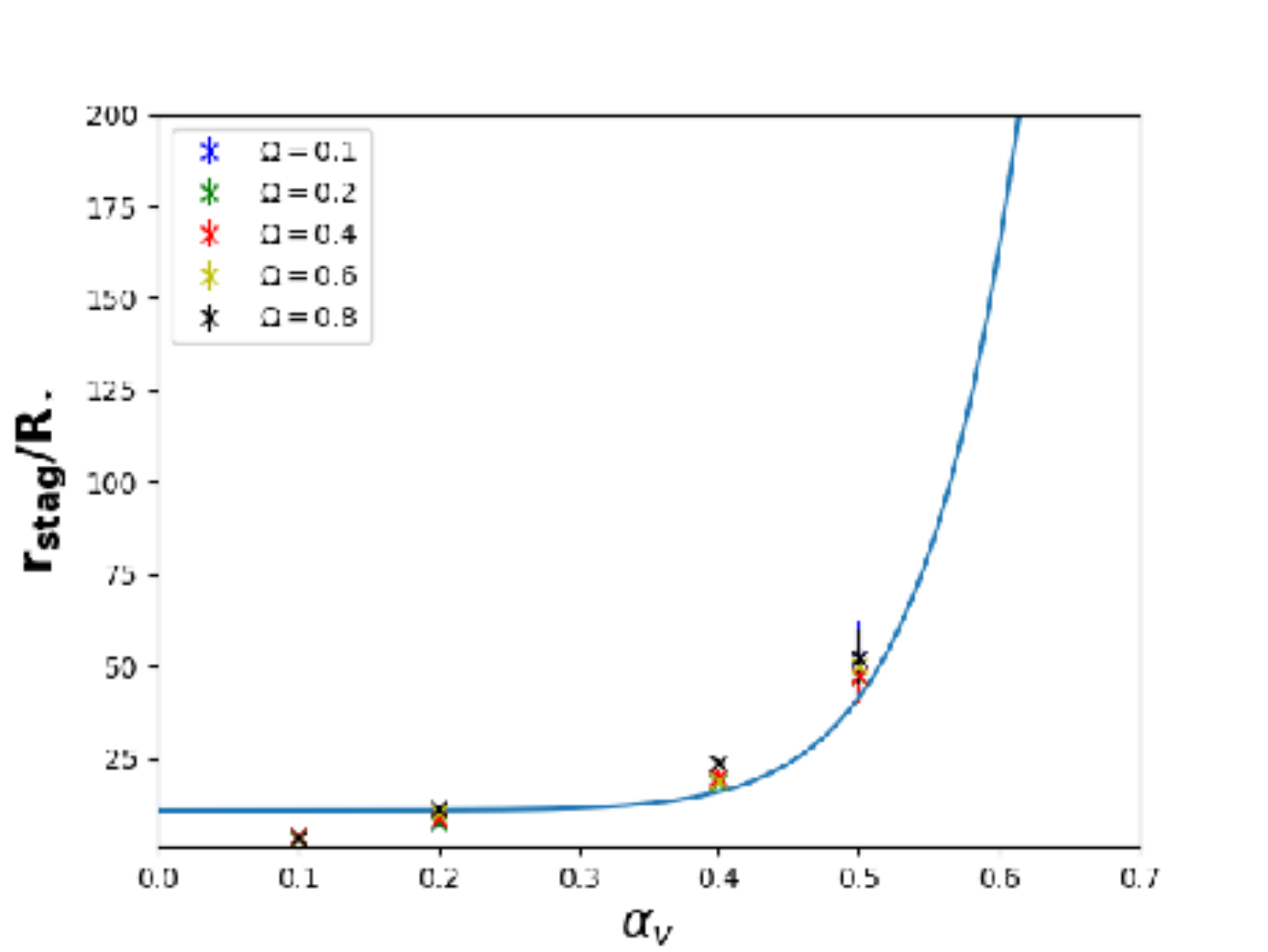}
\caption{Position of the stagnation radius $r_{stag}$ in simulations with
the different viscosity coefficients $\alpha$. The cases
with different rotation rates are represented with the red, blue and
green symbols for simulations with $\Omega_\ast$=0.1, 0.2, 0.4 and
0.6, respectively, with shown estimate of the error bars in positions.
}
\label{rstag}
\end{figure}

\subsection{Backflow in the HD disk}
Purely HD simulations are obtained for different stellar rotation rates
$\Omega$ (scaled with Keplerian break-up velocity) and $\alpha$. For
all the values $\alpha<0.6$, we find backflow in our simulation. The
stagnation radius for different $\alpha$ is marked $r_{stag}$. When
$\alpha >0.6$, there is no backflow in the disk. In the Fig.\ref{backflshd}
we have presented our simulations with different $\alpha$, indicating the
stagnation radius. 

We find that stagnation radius is a function of $\alpha$. It follows
the same trend as predicted by analytical model in KK00.

\section{Conclusions}
We present results of HD and MHD simulations of thin accretion disks with backflow.
The result from numerical simulations is in agreement with the analytical solution:
with viscous coefficient $\alpha<0.6$, backflow occurs in the mid-plane of the
disk. The starting point of backflow, stagnation radius, is a function of $\alpha$,
as predicted by analytical model. For future work we will extend our study
to magnetic cases.

\acknowledgements{This project was funded by a Polish NCN grant no.
2013/08/A/ST9/00795. M\v{C} acknowledges collaboration with Croatian
project STARDUST through HRZZ grant IP-2014-09-8656, and ANR Toupies
grant with A.S. Brun under which he developed the setup for star-disk
simulations while in CEA, Saclay. We thank ASIAA/TIARA (PL and XL
clusters) in Taipei, Taiwan and NCAC (PSK and CHUCK clusters) in
Warsaw, Poland, for access to Linux computer clusters used for
high-performance computations. We thank the {\sc pluto} team for
the possibility to use the code.}

\bibliographystyle{ptapap}
\bibliography{ruchiPTA2019}

\end{document}